\RequirePackage{fix-cm}

\documentclass[smallextended]{svjour3}       % onecolumn (second format)
\smartqed  % flush right qed marks, e.g. at end of proof
\usepackage{amsmath,amssymb}
\usepackage{epstopdf}
\usepackage{graphicx}
\usepackage{dcolumn} % Align table columns on decimal point
\usepackage{bm}      % bold math
\usepackage{subfigure}
\usepackage{color}
\usepackage{url}
\usepackage{booktabs}
\usepackage{multirow}
\usepackage{cite}
\renewcommand\refname{Reference}
\journalname{quantum information processing}
\begin{document}

\title{Three-state quantum walk on the Cayley Graph of the Dihedral Group }

\author{Ying Liu \and
        Jia-bin Yuan \and
        Wen-jing Dai \and
        Dan Li
        }

\institute{Ying Liu \and
        Jia-bin Yuan \and
        Wen-jing Dai \and
        Dan Li \at
         College of Computer Science and Technology, Nanjing University of Aeronautics and Astronautics, Nanjing, 211106, China\\
         \email{liuycherish@sina.com.}
         \and
          Ying Liu \at
           State Key Laboratory of Cryptology, P.O.Box 5159, Beijing, 100878, China.\\         %  \\
}
\date{Received: date / Accepted: date}

\maketitle

\begin{abstract}
The finite dihedral group generated by one rotation and one reflection is the simplest case of the non-abelian group. Cayley graphs are diagrammatic counterparts of groups. In this paper, much attention is given to the Cayley graph of the dihedral group. Considering the characteristics of the elements in the dihedral group, we propose a model of three-state discrete-time quantum walk (DTQW) on the Caylay graph of the dihedral group with Grover coin. We derive analytic expressions for the the position probability distribution and the long-time limit of the return probability starting from the origin. It is shown that the localization effect is governed by  the size of the underlying dihedral group, coin operator and initial state. We also numerically investigate the properties of the proposed model via the  probability distribution and the time-averaged probability at the designated position. The abundant phenomena of three-state Grover DTQW on the Caylay graph of the dihedral group can help the community to better understand and to develop new quantum algorithms.
% \PACS{PACS code1 \and PACS code2 \and more}
% \subclass{MSC code1 \and MSC code2 \and more}
\end{abstract}

\section{Introduction}
\label{sec:1}
Discrete-time quantum walks (DTQWs) \cite{aharonov1993quantum}\cite{aharonov2001quantum}\cite{Ying2018Quantum} have attracted an increasing interest in the past two decades (for reviews, see \cite{venegas2012quantum}\cite{wang2013physical}\cite{portugal2013quantum}), which are quantum analogs of discrete-time classical random walks. Due to their inherent nonlinear chaotic dynamic behavior and quantum interference effects, most of the existing quantum walk algorithms are superior to their
classical counterparts at executing certain computational tasks, e.g., element distinctness \cite{ambainis2007quantum}\cite{belovs2012learning}, triangle finding \cite{magniez2007quantum}\cite{lee2013improved}, verifying matrix products \cite{buhrman2006quantum}, searching for a marked element \cite{shenvi2003quantum}, quantized Google's PageRank \cite{paparo2012google} and graph isomorphism \cite{douglas2008classical}.

During the study of DTQW, one-dimensional two-state DTQW has been extensively studied. Many kinds of models of quantum walks have been proposed, such as multiple coins \cite{brun2003quantum}, multiple walkers \cite{xue2012two}, time-dependent coin \cite{banuls2006quantum}, quantun walk with memory \cite{Li2016Generic}. As an important extension, one-dimensional three-state DTQW was first considered by Inui et al. \cite{inui2005one}, reported localization around an initial position. This phenomenon is previously found in two-state DTQWs on square lattices\cite{inui2004localization}. Since then, the three-state DTQW on the line were examined theoretically and numerically. Researches show that the localization effect happens with a broad family of coin operators in three-state DTQWs \cite{vstefavnak2014limit}\cite{machida2015localization}. Moreover, a weak limit theorem is recently derived in \cite{machida2015limit}\cite{falkner2014weak} for arbitrary coin initial state and coin operator.

Due to its wealth of symmetries, the dihedral group has been studied extensively. They are of particular interest in various fields of mathematics \cite{borel2006seminar}\cite{golubitsky1986hopf}\cite{chattopadhyay2018connectivity}, computer science \cite{kuperberg2005subexponential} and the natural sciences. In particular, we need establish mathematical models for physical systems \cite{hamermesh2012group}\cite{ko2007string} and molecular orbitals \cite{cotton1988advanced}\cite{lomont2014applications} in the natural sciences. In the quantum information context, Kuperberg \cite{kuperberg2005subexponential} presented the first subexponential time algorithm for the dihedral HSP. Namely, his algorithm runs in time  $2^{o\left ( \sqrt{log N} \right )}$ (the input size is $O\left ( logN \right )$ ). However, in order to achieve this running time, Kuperberg¡¯s algorithm requires $2^{o\left ( \sqrt{log N} \right )}$ space. Then Regev \cite{regev2004subexponential} presented an algorithm that requires only polynomial space, i.e. $poly(\log N)$. The running time of our algorithm is still subexponential and only slightly higher than Kuperberg¡¯s algorithm. Carignan \cite{carignan2015characterizing} described a protocol that extracts the average fidelity of the error arising over a group of single-qubit operations corresponding to the dihedral group.

Random walks on groups play an essential role in various fields of natural science, ranging from  solid-state physics, polymer chemistry, and biology to mathematics and computer science. Motivated by the immense success of random walk methods in the design of classical algorithms, we consider the DTQW on the Cayley graph of the dihedral group, previously considered by Dai et al. \cite{dai2018discrete-time}. Here, we will further study the three-state DTQW on the Cayley graph of the dihedral group. In this paper, we present a model of three-state DTQW on the Caylay graph of the dihedral group with Grover coin using  both analytical and numerical methods. We analyze the three-state Grover DTQW on the Caylay graph of the dihedral group  and prove that  the size of the underlying dihedral group and the coin operation in itself can determine whether localization occur. We calculate the long-time limit of the return probability starting from the origin  and discuss its dependence on the initial state and the system size. We anticipate that the abundant phenomena of three-state Grover DTQW on the Caylay graph of the dihedral group can help the community to better understand and to develop new quantum algorithms.

The rest of the paper is organized as follows. Necessary preliminaries for modeling and the model of three-state DTQW with Grover coin on the Cayley graph of the dihedral group are formally defined in Section \ref{sec:2}. The spectral analysis of its evolution operator using Fourier transformation is given in Section \ref{sec:3}. And a time-averaged probability of finding the particle is also introduced in Section \ref{sec:3}, we prove the probability of finding the particle at a fixed itself converges to a nonzero value after infinite long time. Results of numerical simulation are presented in Section \ref{sec:4}. Finally, a short conclusion is given in Section \ref{sec:5}.

\section{The model of the three-state  DTQW}
\label{sec:2}
In this section,  we first briefly  review the notion of the dihedral group and its Cayley graph. Then we present the model of three-state DTQW on the Cayley graph of the dihedral group.
\subsection{Dihedral group and Cayley graph}
\label{sec:2.1}
\paragraph{Dihedral group} The dihedral group ${D_N}$ is a symmetric group of $N$-gons ($N \ge 3$). Let $\sigma$ denote the rotation of the regular $N$-polygon by an angle of ${{2\pi } \mathord{\left/ {\vphantom {{2\pi } N}} \right. \kern-\nulldelimiterspace} N}$ degrees. Let $\tau $ denote the reflection of the regular $N$-polygon around an axis of symmetry. It is isomorphic to the abstract group generated by the element $\sigma$  of order $N$ and the element  $\tau $ of order $2$ subject to the  relation $\sigma \tau  = {\tau \sigma ^{ - 1}}$. That is, ${D_N} =  < \sigma ,\tau |{\sigma ^N} = {\tau ^2} = {1_{{D_N}}},\tau \sigma \tau  = {\sigma ^{ - 1}} > $, where $\sigma$ and  $\tau$ are two generators, and $1_{{D_N}}$ is the identity element of $D_N$.
The $2N$ elements of ${D_N}$  can be written as  $\{ e,{\rm{ }}\sigma ,{\rm{ }}{\sigma ^2},\ldots ,{\rm{ }}{\sigma ^{(N - 1)}},\tau ,{\rm{ }}\tau \sigma ,{\rm{ }}\tau {\sigma ^2},\ldots ,{\rm{ }}\tau {\sigma ^{(N - 1)}}\}$. The first $N$ listed elements are rotations and the remaining $N$ elements are axis-reflections (all of which have order 2).

\paragraph{Semi-direct product} Let $G$ be a group and $N \triangleleft G$ (normal subgroup), $H < G$ (proper subgroup). If $G = NH$ and $N \cap H = \left\{ e \right\}$, where $e$ is the identity of $G$, then $G$ is called a semi-direct product of $N$ and $H$, denoted by ${G} \cong {N}\rtimes{H}$.

The dihedral group ${D_N}$ is isomorphic to the semi-direct product of ${Z_N}$ and ${Z_2}$, denoted by ${D_N} \cong {Z_N}\rtimes{Z_2}$. Each element of the dihedral group can be expressed as ${\tau ^s}{\sigma ^t}$, written as a pair $(s, t)$, where $s \in {Z_2},t \in {Z_N}$ and ${Z_N} = \left\{ {0,1, \cdots ,N - 1} \right\}$. If $s = 0$, $(s, t)$ is called a rotation of the dihedral group. If $s = 1$, $(s, t)$ is called a reflection of the dihedral group.

\paragraph{Cayley graph}
Let $G$ be a finite group, and let $H = \left\{ {{h_1}, \cdots ,{h_k}} \right\}$ be a generating set for $G$. The Cayley graph of $G$ with respect to $H$ has a vertex for every element of $G$, with an oriented edge from $g$ to $gh$, where $\forall g \in G$ and $\forall h \in H$. A Cayley graph of the group $D_N$ can be derived from the group presentation ${D_N} =  < \sigma ,\tau |{\sigma ^N} = {\tau ^2} = {1_{{D_N}}},\tau \sigma \tau  = {\sigma ^{ - 1}} > $. The graph is mixed: it has $2N$ vertices, $2N$ arrows, and $N$ edges.

\subsection{Time evolution of the three-state DTQW }
\label{sec:2.2}
In this part, we present the model of three-state DTQW on the Cayley graph of the dihedral group. The generators of dihedral group have not only rotation and reflection operation, but also identity operation. So we add a new direction -staying at the same position. That is to say, the walker governed by three-state quantum walk moves to the rotation, the reflection, and stays at the same position.

\begin{figure}
\centering
\includegraphics[height=4.5cm]{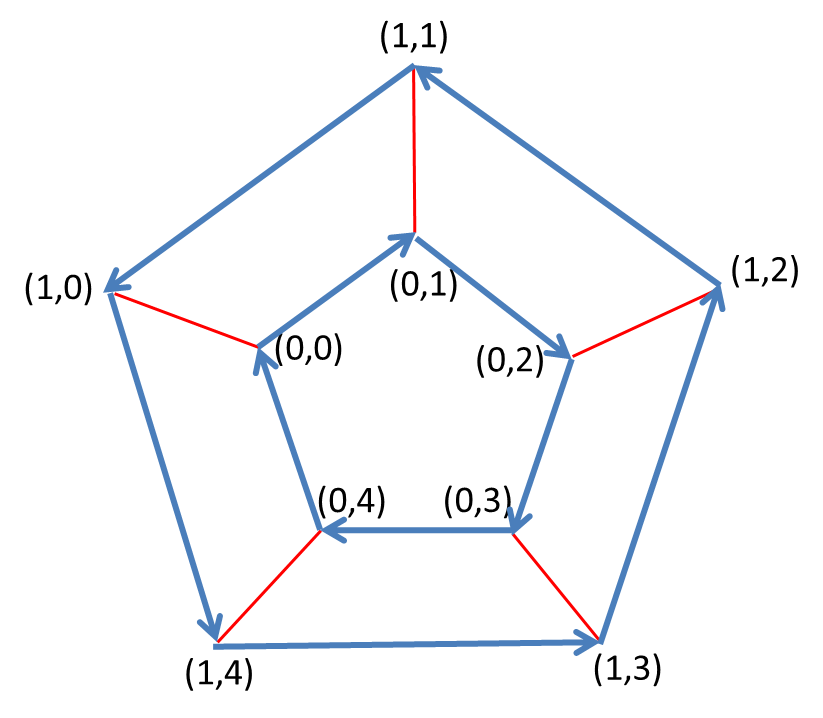}
\caption{Good encoding of the Cayley graph of the dihedral group $D_5$.}
\label{fig:1}
\end{figure}

The dihedral group consists of  $N$ rotations and $N$ reflections, which we denote by $\left( {0,{\rm{ }}0} \right),{\rm{ }}.{\rm{ }}.{\rm{ }}.{\rm{ }},\left( {0,{\rm{ }}N{\rm{ }} - {\rm{ }}1} \right)$ and $\left( {1,0} \right),{\rm{ }}.{\rm{ }}.{\rm{ }}.{\rm{ }},\left( {1,{\rm{ }}N{\rm{ }} - {\rm{ }}1} \right)$ respectively. Each element of the dihedral group is denoted by a pair $(s, t)$, thus vertices of the Cayley graph of the dihedral group can be encoded as a pair $(s, t)$. Fig.\ref{fig:1} shows the example encoding of the Cayley graph of the dihedral group $D_5$. If the pair $(s, t)$ corresponds to two registers $\left| s \right\rangle \left| t \right\rangle $, the vertex set is denoted by
\begin{equation}
V = \left\{ {\left| 0 \right\rangle \left| 0 \right\rangle ,\left| 0 \right\rangle \left| 1 \right\rangle,  \cdots ,\left| 0 \right\rangle \left| {N - 1} \right\rangle ,\left| 1 \right\rangle \left| 0 \right\rangle ,\left| 1 \right\rangle \left| 1 \right\rangle,  \cdots ,\left| 1 \right\rangle \left| {N - 1} \right\rangle } \right\}.
\end{equation}

\begin{figure}
\centering
\includegraphics[height=4.5cm]{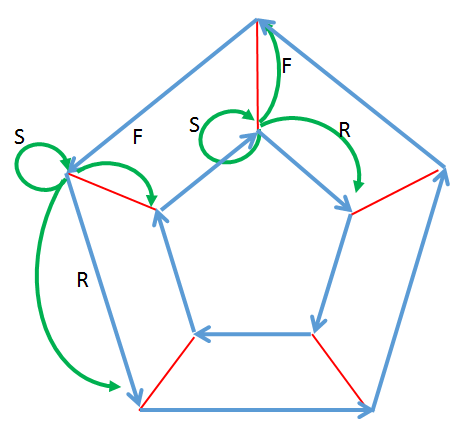}
\caption{The DTQW with a three dimensional coin on the Cayley graph of the dihedral group $D_5$. $R$, $F$, and $S$ facilitate a rotation hop, a reflection hop, or stay put, respectively.}
\label{fig:2}
\end{figure}

The whole Hilbert space of the model is depicted as $H = {H_c} \otimes {H_p}$, where ${H_p}$  is called a  position  Hilbert space which is spanned by an orthogonal normal$\{ \left| {s,j} \right\rangle :s \in \{ 0,1\} ,j \in \{ 0,1, \cdots ,N - 1\} \}$ and ${H_c}$ is  the coin Hilbert space which is spanned by an orthogonal normal basis$\{ \left| 0 \right\rangle ,\left| 1 \right\rangle ,\left| 2 \right\rangle \} $.Considering the characteristics of the
elements in the dihedral group, the particle has three directions, a rotation hop $R$, stay put $S$, and a reflection hop $F$, $R$, $S$, $F$ in each step, for three-state DTQW on the Cayley graph of the dihedral group. The shift operator allows the walker to go one step rotation if the accompanying coin state is ${\left| 0 \right\rangle }$ , one step stay put if the accompanying coin state is  $\left| 1 \right\rangle $, and one step reflection if the accompanying coin state is  $\left| 2 \right\rangle $, as shown in Fig.\ref{fig:2}.

The evolution of the whole system at each step of the walk can be described by the global unitary operator, denoted by $U$
 \begin{equation}
 \begin{array}{c}
 U = S(C \otimes {I_P}),
 \end{array}
\end{equation}
where  ${I_p}$ is the identity of ${H_p}$, $S$  is the shift operator,expressed as
 \begin{equation}
 \begin{array}{c}
S = \left| 0 \right\rangle \left. {\langle 0} \right| \otimes \left| 0 \right\rangle \left. {\langle 0} \right| \otimes \sum\limits_j {\left| {j + 1\left( {\bmod N} \right)} \right\rangle \left. {\langle j} \right|} \\
  + \left| 0 \right\rangle \left. {\langle 0} \right| \otimes \left| 1 \right\rangle \left. {\langle 1} \right| \otimes \sum\limits_j {\left| {j - 1\left( {\bmod N} \right)} \right\rangle \left. {\langle j} \right|} \\
{\rm{ + }}\left| 1 \right\rangle \left. {\langle 1} \right| \otimes \left| 0 \right\rangle \left. {\langle 0} \right| \otimes \sum\limits_j {\left| {j\left( {\bmod N} \right)} \right\rangle \left. {\langle j} \right|} \\
{\rm{ + }}\left| 1 \right\rangle \left. {\langle 1} \right| \otimes \left| 1 \right\rangle \left. {\langle 1} \right| \otimes \sum\limits_j {\left| {j\left( {\bmod N} \right)} \right\rangle \left. {\langle j} \right|} \\
 + \left| 2 \right\rangle \left. {\langle 2} \right| \otimes \left| 0 \right\rangle \left\langle 1 \right| \otimes \sum\limits_j {\left| {j\left( {\bmod N} \right)} \right\rangle \left. {\langle j} \right|} \\
 + \left| 2 \right\rangle \left. {\langle 2} \right| \otimes \left| 1 \right\rangle \left\langle 0 \right| \otimes \sum\limits_j {\left| {j\left( {\bmod N} \right)} \right\rangle \left\langle j \right|}.
\end{array}
\end{equation}
 while $C$ is the coin operation. In this paper, we choose Grover operator as the coin operator, the three-dimensional Grover operator is defined as
\begin{equation}
C = \frac{1}{3}\left( {\begin{array}{*{20}{c}}
{ - 1}&2&2\\
2&{ - 1}&2\\
2&2&{ - 1}
\end{array}} \right).
\end{equation}

Given a initial state $\left| {\Psi (0)} \right\rangle$, the state of the particle after $t$ steps is given by the successive application of the unitary propagator $U$ on the initial state
\begin{equation}
{\left| {\Psi (t)} \right\rangle  = {U^t}\left| {\Psi (0)} \right\rangle}.
\end{equation}
In this paper, we assume the walker always starts at position $\left| {00} \right\rangle $ and the initial coin state satisfies $\Psi (0,0) = \alpha \left| 0 \right\rangle  + \beta \left| 1 \right\rangle  + \gamma \left| 2 \right\rangle $ where $\alpha ,\beta ,\gamma  \in \bm{C}$, ${\left |\alpha \right |^2} + {\left |\beta \right |^2} + {\left |\gamma \right |^2} = 1$. We guarantee the quantum walker starts with its coin state in superposition of only rotation, reflection, and no change in position bases. Therefore, the system's initial state can be formulated as
\begin{equation}
\Psi (0) = \left| {00} \right\rangle  \otimes [\alpha \left| 0 \right\rangle  + \beta \left| 1 \right\rangle  + \gamma \left| 2 \right\rangle ].
\label{eq1}
\end{equation}

\section{ Fourier Analysis}
\label{sec:3}
Fourier transformation is a powerful technique for analysis of classical random walks. For the more, Fourier analysis generalises well to quantum walks, which was introduced to quantum walks by Grimmett et al \cite{Grimmett2004Weak}. In this section, we use Fourier analysis to analytically study the return probability of the three-state quantum walk as $t \to \infty$.
\subsection{Wave function of the three-state  DTQW}
 \label{sec:3.1}
We now define the state of the walker, at time ${\rm{t}}$  and position $x=(s,x)$ £¨$s \in \{ 0,1\} $, ${\rm{x}} \in N$£© to be a 6-dimensional vector. We denote this as
\begin{equation}
\Psi (x,t) = \left[ {\begin{array}{*{20}{c}}
{\left\langle {0,0,x\left| {\Psi (t)} \right.} \right\rangle }\\
{\left\langle {0,1,x\left| {\Psi (t)} \right.} \right\rangle }\\
{\left\langle {1,0,x\left| {\Psi (t)} \right.} \right\rangle }\\
\begin{array}{l}
\left\langle {1,1,x\left| {\Psi (t)} \right.} \right\rangle \\
\left\langle {2,0,x\left| {\Psi (t)} \right.} \right\rangle \\
\left\langle {2,1,x\left| {\Psi (t)} \right.} \right\rangle
\end{array}
\end{array}} \right].
\end{equation}
Expanding $\left| {\Psi (t+1)} \right\rangle  = U\left| {\Psi (t)} \right\rangle$ in terms of $\Psi (x,t)$, we obtain the master equation for the walker at position $x$

\begin{equation}
\Psi (x,t + 1) = {M_1}\Psi (x + 1,t) + {M_2}\Psi (x - 1,t) + {M_3}\Psi (x,t),
\label{eq2}
\end{equation}
where ${M_1}$, ${M_2}$ and ${M_3}$ matrices are
$
{M_1} = \frac{1}{{3}}\left[ {\begin{array}{*{20}{c}}
0&0&0&0&0&0\\
0&-1&0&2&0&2\\
0&0&0&0&0&0\\
0&0&0&0&0&0\\
0&0&0&0&0&0\\
0&0&0&0&0&0
\end{array}} \right]
$,
$
{M_2} = \frac{1}{{3}}\left[ {\begin{array}{*{20}{c}}
-1&0&2&0&2&0\\
0&0&0&0&0&0\\
0&0&0&0&0&0\\
0&0&0&0&0&0\\
0&0&0&0&0&0\\
0&0&0&0&0&0
\end{array}} \right]
$,
$
{M_3} = \frac{1}{{3}}\left[ {\begin{array}{*{20}{c}}
0&0&0&0&0&0\\
0&0&0&0&0&0\\
2&0&-1&0&2&0\\
0&2&0&-1&0&2\\
0&2&0&2&0&-1\\
2&0&2&0&-1&0
\end{array}} \right]
$.

Our objective is to find analytical expression for $\Psi (x,t)$. To do so, we compute the Discrete Time Fourier Transform of Eq.(\ref{eq2}). The Discrete Time Fourier Transform on the system state is given by
\begin{equation}
\tilde \Psi (k,t) = \sum\limits_{x = 0}^{N - 1} {{e^{2\pi ikx/N}}} \Psi (x,t),k \in \{ 0,1, \cdots ,N-1\}.
\label{eq3}
\end{equation}
The corresponding inverse Fourier transform is
\begin{equation}
\Psi (x,t) = \sum\limits_{k = 0}^{N - 1} {{e^{ - 2\pi ikx/N}}} \tilde \Psi (k,t).
\end{equation}

From now on, a tilde indicates quantities with a $k$ dependence. Applying the Fourier transformation Eq.(\ref{eq3}) to the time evolution equation Eq.(\ref{eq2}), after some algebra we get the master equation in Fourier space:
\begin{equation}
 {\tilde \Psi (k,t + 1)}  = \underbrace {({M_1}{e^{ - 2\pi ik/N}} + {M_2}{e^{2\pi ik/N}} + {M_3})}_{M_k}{\tilde \Psi (k,t)}.
 \label{eq4}
\end{equation}
Let $K = \omega _N^{ k} = {e^{ \frac{{2\pi ik}}{N}}}$, ${M_k}$ has the form of
\begin{equation}
{M_k} = \frac{1}{3}\left[ {\begin{array}{*{20}{c}}
{ - K}&0&{2K}&0&{2K}&0\\
0&{{K^{ - 1}}}&0&{2{K^{ - 1}}}&0&{2{K^{ - 1}}}\\
2&0&{ - 1}&0&2&0\\
0&2&0&{ - 1}&0&2\\
0&2&0&2&0&{ - 1}\\
2&0&2&0&{ - 1}&0
\end{array}} \right].
\end{equation}
To express the wave function as a function of $t$ explicitly we consider the eigenvalues and  eigenvectors of $M_k$. Let ${\lambda _j}(k)$ $(j = 1,2, \cdots ,6)$ be the eigenvalues of $M_k$ and the corresponding eigenvector are ${{\nu _j}(k)}$. After some calculations, we get explicit forms of the eigenvalues
\begin{equation}
\begin{array}{*{20}{c}}
{\lambda _1}(k) = -1,
{\lambda _2}(k) = 1,\\
{\lambda _3}(k) = \frac{{1 - 2\sqrt 2 i}}{3},

{\lambda _4}(k) = \frac{{1 + 2\sqrt 2 i}}{3},\\
{\lambda _5}(k) = -\frac{{A }}{{6K}},
{\lambda _6}(k) = -\frac{{B }}{{6K}}.
\end{array}
\end{equation}
The corresponding eigenvectors are
\begin{equation}
\begin{array}{*{20}{c}}
\left| {{\nu _1}(k)} \right\rangle  = \left[ {\begin{array}{*{20}{c}}
0\\
0\\
{ - 1}\\
{ - 1}\\
1\\
1
\end{array}} \right],
\left| {{\nu _2}(k)} \right\rangle  = \left[ {\begin{array}{*{20}{c}}
1\\
{{K^{ - 1}}}\\
{\frac{{K + 1}}{{2K}}}\\
{\frac{{K + 1}}{{2K}}}\\
{{K^{ - 1}}}\\
1
\end{array}} \right],\\
\left| {{\nu _3}(k)} \right\rangle  = \left[ {\begin{array}{*{20}{c}}
{ - \frac{{2\sqrt 2 iK}}{{3K + 1}}}\\
{\frac{{2\sqrt 2 i}}{{3K + 1}}}\\
{\frac{{K - 1 - \sqrt 2 iK - \sqrt 2 i}}{{3K + 1}}}\\
{\frac{{K - 1 + \sqrt 2 iK + \sqrt 2 i}}{{3K + 1}}}\\
{ - \frac{{K + 3}}{{3K + 1}}}\\
1
\end{array}} \right],
\left| {{\nu _4}(k)} \right\rangle  = \left[ {\begin{array}{*{20}{c}}
{\frac{{2\sqrt 2 iK}}{{3K + 1}}}\\
{ - \frac{{2\sqrt 2 i}}{{3K + 1}}}\\
{\frac{{K - 1 + \sqrt 2 iK + \sqrt 2 i}}{{3K + 1}}}\\
{\frac{{K - 1 - \sqrt 2 iK - \sqrt 2 i}}{{3K + 1}}}\\
{ - \frac{{K + 3}}{{3K + 1}}}\\
1
\end{array}} \right],\\
\left| {{\nu _5}(k)} \right\rangle  = \left[ {\begin{array}{*{20}{c}}
{\frac{{5K + 1}}{{2K(K - 1)}} - \frac{{(K + 1)A}}{{4K(K - 1)}}}\\
{ - \frac{{K + 5}}{{2K - 1}} + \frac{{(K + 1)A}}{{4{K^2}(K - 1)}}}\\
{ - \frac{{2K + 1}}{{K(K - 1)}} + \frac{A}{{2K(K - 1)}}}\\
{\frac{{K + 2}}{{K - 1}} - \frac{A}{{2K(K - 1)}}}\\
{{K^{ - 1}}}\\
1
\end{array}} \right],
\left| {{\nu _6}(k)} \right\rangle  = \left[ {\begin{array}{*{20}{c}}
{\frac{{5K + 1}}{{2K(K - 1)}} - \frac{{(K + 1)B}}{{4K(K - 1)}}}\\
{ - \frac{{K + 5}}{{2K - 1}} + \frac{{(K + 1)B}}{{4{K^2}(K - 1)}}}\\
{ - \frac{{2K + 1}}{{K(K - 1)}} + \frac{B}{{2K(K - 1)}}}\\
{\frac{{K + 2}}{{K - 1}} - \frac{B}{{2K(K - 1)}}}\\
{{K^{ - 1}}}\\
1
\end{array}} \right],
\end{array}
\end{equation}
where $A = {K^2} + 4K + 1 - K\sqrt {{K^2} + 10K + 1}  + \sqrt {{K^2} + 10K + 1} $ and $B = {K^2} + 4K + 1 + K\sqrt {{K^2} + 10K + 1}  - \sqrt {{K^2} + 10K + 1} $.

 Puting $M_k$ in its eigenbasis, then using the mathematical properties of linear operators, we can rewrite Eq.(\ref{eq4}) as
\begin{equation}
\begin{array}{c}
\left| {\tilde \Psi (k,t)} \right\rangle  = M_k\left| {\tilde \Psi (k,t-1)} \right\rangle  = M_k^t\left| {\tilde \Psi (k,0)} \right\rangle \\
 = \sum\limits_{j = 1}^6 {\lambda _j^t(k)\left\langle {{{\omega _j}(k)}}
 \mathrel{\left | {\vphantom {{{\omega _j}(k)} {\tilde \Psi (k,0)}}}
 \right. \kern-\nulldelimiterspace}
 {{\tilde \Psi (k,0)}} \right\rangle \left| {{\omega _j}(k)} \right\rangle}.
\end{array}
\end{equation}
$\tilde \Psi (k,0)$ is the Fourier transformed system's initial state, by Eq.(\ref{eq3}) we get $\tilde \Psi (k,0) = \mathbf{\left[ {\alpha ,\beta ,\gamma ,0, \cdots ,0} \right]}^\mathsf{T}$, and
\begin{equation}
\left| {{\omega _j}(k)} \right\rangle  = \frac{1}{{\sqrt {{N_j}(k)} }}\left| {{\nu _j}(k)} \right\rangle,
\label{eq5}
\end{equation}
where ${{N_j}(k)}$ is a normalized factor. Then the wave function is obtained by the inverse Fourier transform:

\begin{equation}
\begin{array}{c}
\left| {\Psi (x,t)} \right\rangle  = \frac{1}{N}\sum\limits_{k = 0}^{N - 1} {{e^{{\raise0.7ex\hbox{${2\pi ikx}$} \!\mathord{\left/
 {\vphantom {{2\pi ikx} N}}\right.\kern-\nulldelimiterspace}
\!\lower0.7ex\hbox{$N$}}}}\left| {\tilde \Psi (k,t)} \right\rangle } \\
 = \frac{1}{N}\sum\limits_{k = 0}^{N - 1} {\sum\limits_{j = 1}^6 {{e^{{\raise0.7ex\hbox{${2\pi ikx}$} \!\mathord{\left/
 {\vphantom {{2\pi ikx} N}}\right.\kern-\nulldelimiterspace}
\!\lower0.7ex\hbox{$N$}}}}} } \lambda _j^t(k)\left\langle {{{\omega _j}(k)}}
 \mathrel{\left | {\vphantom {{{\omega _j}(k)} {\Psi (k,0)}}}
 \right. \kern-\nulldelimiterspace}
 {{\tilde \Psi (k,0)}} \right\rangle \left| {{\omega _j}(k)} \right\rangle.
\end{array}
\end{equation}

In the above calculations, it is most important to emphasize that the eigenvalue $1$ and $-1$ independent of the value $k$ exist in the three-state Grover DTQW. This distinctive property causes the particular time evolution for the three-state Grover DTQW. Inui\cite{inui2004localization} has shown that the existence of a strongly degenerated eigenvalue such as $1$ or $-1$ is a necessary condition for the quantum walk showing the localization. The number of the degenerate eigenvalues of the time evolution matrix has a profound effect on quantum walks showing the localization.

\subsection{Limit theorem for the three-state DTQW starting from the origin}
\label{sec:3.2}
From now on we compute the time-averaged probability at the origin in the three-state Grover DTQW on the Cayley graph of the dihedral group with $N$ vertices. We begin with considering the probability $P({X_t} = x)$ that the walker can be found at position $x$  and at time $t$ starting from an initial state $\left| {\Psi (0)} \right\rangle$. The probability $P({X_t} = x)$ is calculated from the relation
\begin{equation}
\begin{aligned}
P({X_t} = x)& = \left\langle {{\Psi (x,t)}}
 \mathrel{\left | {\vphantom {{\Psi (x,t)} {\Psi (x,t)}}}
 \right. \kern-\nulldelimiterspace}
 {{\Psi (x,t)}} \right\rangle \\
 &= \frac{1}{{{N^2}}}\sum\limits_{k,m = 0}^{N - 1} {\sum\limits_{j,l = 1}^6 {{e^{{\raise0.7ex\hbox{${2\pi i(m - k)x}$} \!\mathord{\left/
 {\vphantom {{2\pi i(m - k)x} N}}\right.\kern-\nulldelimiterspace}
\!\lower0.7ex\hbox{$N$}}}}} } {({\lambda _j}{(k)^ * }{\lambda _l}(m))^t}\\&{(\left\langle {{{\omega _j}(k)}}
 \mathrel{\left | {\vphantom {{{\omega _j}(k)} {\tilde \Psi (k,0)}}}
 \right. \kern-\nulldelimiterspace}
 {{\tilde \Psi (k,0)}} \right\rangle )^ * }\left\langle {{{\omega _l}(m)}}
 \mathrel{\left | {\vphantom {{{\omega _l}(m)} {\tilde \Psi (m,0)}}}
 \right. \kern-\nulldelimiterspace}
 {{\tilde \Psi (m,0)}} \right\rangle \left\langle {{{\omega _j}(k)}}
 \mathrel{\left | {\vphantom {{{\omega _j}(k)} {{\omega _l}(m)}}}
 \right. \kern-\nulldelimiterspace}
 {{{\omega _l}(m)}} \right\rangle.
\end{aligned}
\end{equation}
The probability distribution  $P({X_t} = x)$ does not converge in the limit $t \to \infty$, while its average over time does. Thus we introduce time-averaged probability distribution defined by:

\begin{equation}
\begin{aligned}
\bar P({X_t} = x) &= \frac{1}{T}\sum\nolimits_{t = 0}^{T - 1} {P({X_t} = x)}\\&= \frac{1}{T}\sum\nolimits_{t = 0}^{T - 1} {\left\langle {{\Psi (x,t)}}
 \mathrel{\left | {\vphantom {{\Psi (x,t)} {\Psi (x,t)}}}
 \right. \kern-\nulldelimiterspace}
 {{\Psi (x,t)}} \right\rangle }.
\end{aligned}
\label{eq6}
\end{equation}

Next, we focus on the localization phenomenon on the three-state DTQW. To determine whether the localization will occur at the origin, we need to calculate a return probability of the three-state walk as $t \to \infty $. Since the walker starts from the origin, we consider the return probability as the probability that the walker can be observed at the origin. That is, the return probability at time $t$ is determined by the probability $\mathop {\lim }\limits_{t \to \infty } \bar P({X_t} = 0)$. Let the probability be ${\hat P}_0$, where we use a hat to indicate the asymptotic limit of $t$. As $t \to \infty $, we obtain the following limit theorem about the return probability.

\paragraph{Theorem 1} If the walker starts with the state given by Eq.(\ref{eq1}), the return probability is of the form
\begin{equation}
\begin{array}{c}
{{\hat P}_0} = \mathop {\lim }\limits_{t \to \infty } \bar P({X_t} = 0)\\
 = \frac{1}{{{N^2}}}\sum\limits_{k = 0}^{N - 1} {\sum\limits_{j = 1}^6 {{{(\left\langle {{{\omega _j}(k)}}
 \mathrel{\left | {\vphantom {{{\omega _j}(k)} {\tilde \Psi (k,0)}}}
 \right. \kern-\nulldelimiterspace}
 {{\tilde \Psi (k,0)}} \right\rangle )}^ * }(\left\langle {{{\omega _j}(k)}}
 \mathrel{\left | {\vphantom {{{\omega _j}(k)} {\tilde \Psi (k,0)}}}
 \right. \kern-\nulldelimiterspace}
 {{\tilde \Psi (k,0)}} \right\rangle )} },
\end{array}
\end{equation}
where $\tilde \Psi (k,0)$ is the Fourier transformed system's initial state, and $\left| {{\omega _j}(k)} \right\rangle $ is the normalized eigenvectors.

\textbf{Proof: }  We start by obtaining the specific formula of  ${\hat P}_0$  using Eq.(\ref{eq6})
\begin{equation}
\begin{aligned}
{{\hat P}_0}& \equiv \mathop {\lim }\limits_{t \to \infty } \bar P({X_t} = 0)\\
 &= \mathop {\lim }\limits_{T \to \infty } \frac{1}{T}\sum\nolimits_{t = 0}^{T - 1} {\left\langle {{\Psi (0,t)}}
 \mathrel{\left | {\vphantom {{\Psi (0,t)} {\Psi (0,t)}}}
 \right. \kern-\nulldelimiterspace}
 {{\Psi (0,t)}} \right\rangle } \\
 &= \frac{1}{{{N^2}}}\sum\limits_{k,m = 0}^{N - 1} {\sum\limits_{j,l = 1}^6 {{{(\left\langle {{{\omega _j}(k)}}
 \mathrel{\left | {\vphantom {{{\omega _j}(k)} {\tilde \Psi (k,0)}}}
 \right. \kern-\nulldelimiterspace}
 {{\tilde \Psi (k,0)}} \right\rangle )}^ * }\left\langle {{{\omega _l}(m)}}
 \mathrel{\left | {\vphantom {{{\omega _l}(m)} {\tilde \Psi (m,0)}}}
 \right. \kern-\nulldelimiterspace}
 {{\tilde \Psi (m,0)}} \right\rangle \left\langle {{{\omega _j}(k)}}
 \mathrel{\left | {\vphantom {{{\omega _j}(k)} {{\omega _l}(m)}}}
 \right. \kern-\nulldelimiterspace}
 {{{\omega _l}(m)}} \right\rangle } }
 \\& \mathop {\lim }\limits_{T \to \infty } \frac{1}{T}\sum\nolimits_{t = 0}^{T - 1} {{{({\lambda _j}{{(k)}^ * }{\lambda _l}(m))}^t}}.
\end{aligned}
\end{equation}

Obviously, ${{({\lambda _j}{{(k)}^ * }{\lambda _l}(m))}^t}$ is the only time dependent term in the above expression. Therefor, we are interested in
\begin{equation}
\frac{1}{T}\sum\nolimits_{t = 0}^{T - 1} {{{({\lambda _j}{{(k)}^ * }{\lambda _l}(m))}^t}}.
\label{eq7}
\end{equation}
which is divided the following two cases:  the first kind of circumstance is ${\lambda _j}(k) = {\lambda _l}(m)$. In this case, we have that the average
in Eq. (\ref{eq7}) is equal to $1$. The second is ${\lambda _j}(k) \neq  {\lambda _l}(m)$. we can write
\begin{equation}
\frac{1}{T}\sum\nolimits_{t = 0}^{T - 1} {{{({\lambda _j}{{(k)}^ * }{\lambda _l}(m))}^t}}  = \frac{{\left| {1 - {{({\lambda _j}{{(k)}^ * }{\lambda _l}(m))}^T}} \right|}}{{T\left| {1 - {\lambda _j}{{(k)}^ * }{\lambda _l}(m)} \right|}} \le \frac{2}{{T\left| {{\lambda _j}(k) - {\lambda _l}(m)} \right|}}.
\end{equation}
The latter term converges to zero, therefore the contribution to the limiting distribution comes solely from terms with
${\lambda _j}(k) = {\lambda _l}(m)$. We have know that all the eigenvalues of $U$ are different. Hence, ${\lambda _j}(k) = {\lambda _l}(m)$ if and only if $j=l$ and $k=m$. Further, the limit of the return probability at origin ${\hat P}_0$ can be  obtained now\\

$\begin{array}{c}
{{\hat P}_0} = \mathop {\lim }\limits_{t \to \infty } \bar P({X_t} = 0)\\
 = \frac{1}{{{N^2}}}\sum\limits_{k = 0}^{N - 1} {\sum\limits_{j = 1}^6 {{{(\left\langle {{{\omega _j}(k)}}
 \mathrel{\left | {\vphantom {{{\omega _j}(k)} {\tilde \Psi (k,0)}}}
 \right. \kern-\nulldelimiterspace}
 {{\tilde \Psi (k,0)}} \right\rangle )}^ * }(\left\langle {{{\omega _j}(k)}}
 \mathrel{\left | {\vphantom {{{\omega _j}(k)} {\tilde \Psi (k,0)}}}
 \right. \kern-\nulldelimiterspace}
 {{\tilde \Psi (k,0)}} \right\rangle )} }.
\end{array}$

Observing theorem 1, the exact form of ${\hat P}_0$ can be obtained by exploiting the eigenvectors Eq.(\ref{eq5}). It is obvious that the explicit form for ${\hat P}_0$ is a nonzero constant depending on the parameters $\alpha ,\beta ,\gamma$ and the system size $N$. Thus the existence of localization is directly related to the system's initial coin state and the number of vertices.

\section{Numerical Simulation}
\label{sec:4}
In this part, numerical simulation of the three-state DTQW on the Cayley graph of the dihedral group is provided with MATLAB code to study  the basic
properties of the walk. The numerical simulation helps us verify the results of theoretical analysis. We focus on  probability distribution and Time-averaged probability of the three-state DTQW on the Cayley graph of the dihedral group.

In the numerical simulation, the vertices  in the Cayley graph of the dihedral group need to be re-coded. The specific coding is as follow:
\begin{equation}
\begin{array}{c}
V = \{ \left| 0 \right\rangle \left| 0 \right\rangle ,\left| 0 \right\rangle \left| 1 \right\rangle , \cdots ,\left| 0 \right\rangle \left| {N - 1} \right\rangle ,\left| 1 \right\rangle \left| 0 \right\rangle ,\left| 1 \right\rangle \left| 1 \right\rangle , \cdots ,\left| 1 \right\rangle \left| {N - 1} \right\rangle \} \\
 = \{ 0,1, \cdots ,N - 1,N,N + 1, \cdots ,2N - 1\}
\end{array}.
\end{equation}
That is,
${\left| 0 \right\rangle _V}{\left| 0 \right\rangle _V}$ corresponding to $0$,
${\left| 0 \right\rangle _V}{\left| x \right\rangle _V}$ corresponding to $x$,
${\left| 1 \right\rangle _V}{\left| 0 \right\rangle _V}$ corresponding to $N$,
${\left| 1 \right\rangle _V}{\left| x \right\rangle _V}$ corresponding to $N+x$.

\subsection{probability distribution}
\label{sec:4.1}
To begin with, we show the probability distribution of the three-state DTQW on the Cayley graph of the dihedral group, in order to study how the initial state affects the property of the walk.

\begin{figure}
\centering
\subfigure[]{
\includegraphics[height=3.5cm]{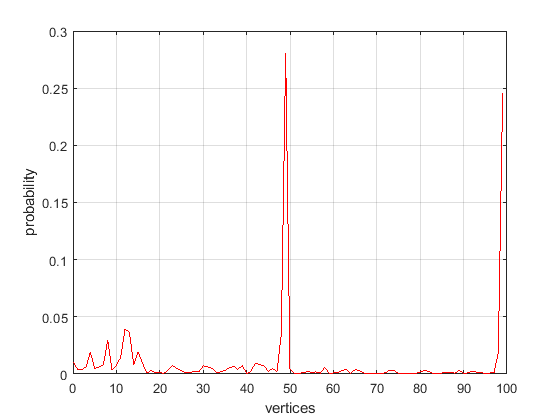}
}
\subfigure[]{
\includegraphics[height=3.5cm]{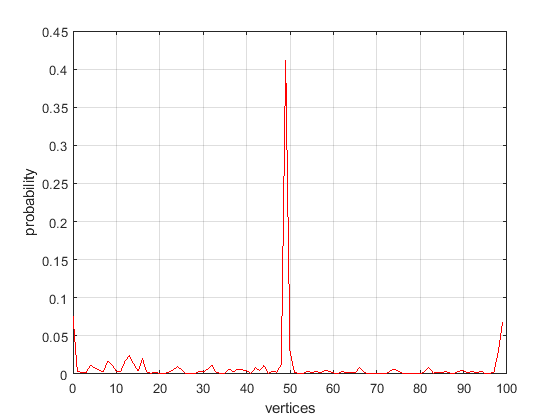}
}
\subfigure[]{
\includegraphics[height=3.5cm]{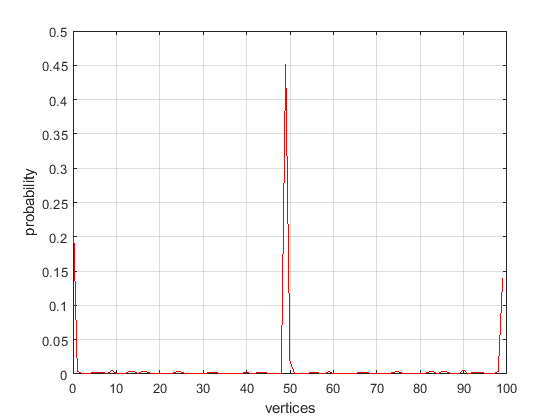}
}
\subfigure[]{
\includegraphics[height=3.5cm]{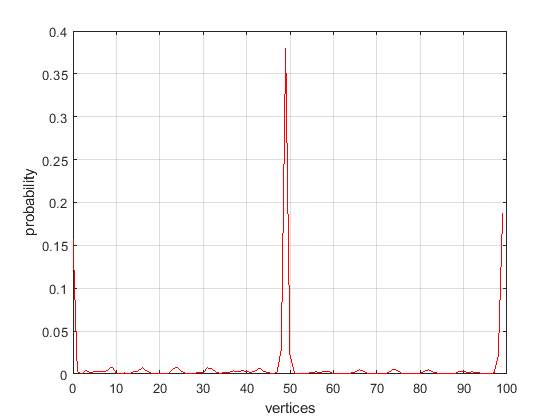}
}
\renewcommand{\figurename}{Fig.}
\caption{Probability distribution of the  three-state Grover DTQW on the Cayley graph of the dihedral group after  200 steps. This is for a dihedral group $D_N$ of size $N = 50$ with various initial coin states:  ${\left| 0 \right\rangle}$ (Plot (a)),  ${\left| 1 \right\rangle}$ (plot (b)),  ${\left| 2 \right\rangle}$ (plot (c)), $\frac{1}{{\sqrt 3 }}(\left| 0 \right\rangle  + \left| 1 \right\rangle  + \left| 2 \right\rangle )$ (plot (d)).}
\label{fig:3}
\end{figure}

 We choose the initial position state  ${\left| 1 0 \right\rangle _V}$ and the initial coin state are respectively ${\left| 0 \right\rangle }$,${\left| 1 \right\rangle }$,${\left| 2 \right\rangle }$ in Hilbert space $H_C^3$, and $\frac{1}{{\sqrt 3 }}(\left| 0 \right\rangle  + \left| 1 \right\rangle  + \left| 2 \right\rangle )$. Fig.\ref{fig:3} illustrates  the numerical simulation results of probability distribution of the three-state DTQW on the Cayley graph of the dihedral group $D_N$ with $N = 50$ after 200 steps. A noticeable peak arises around  the starting vertex and its reflection vertex  in the picture. Whatever the initial coin state is, peak exists around the origin and its reflection vertex in the pictures. The different initial coin states control how much of the distribution is in the central peaks.

\begin{figure}
\centering
\subfigure[]{
\includegraphics[height=3.5cm]{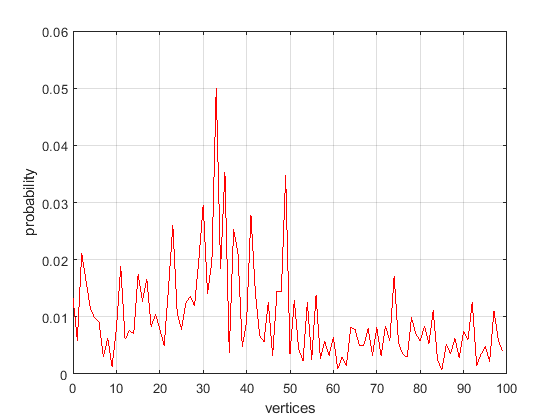}
}
\subfigure[]{
\includegraphics[height=3.5cm]{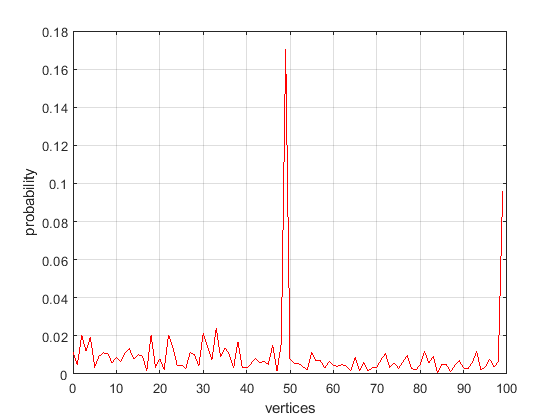}
}
\subfigure[]{
\includegraphics[height=3.5cm]{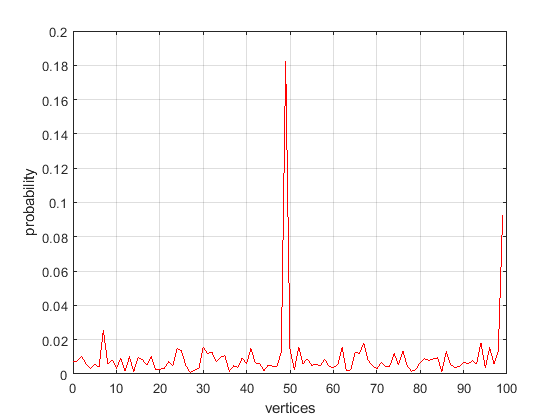}
}
\subfigure[]{
\includegraphics[height=3.5cm]{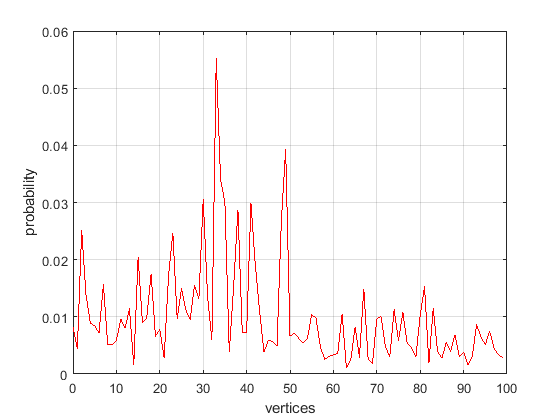}
}
\renewcommand{\figurename}{Fig.}
\caption{As Fig.\ref{fig:3} but with DFT coin.
 }
\label{fig:4}
\end{figure}

The Grover coin is biased but symmetric, while DFT(Discrete Fourier Transform) coin is unbiased but asymmetric. The degree-3 DFT coin is defined as
\begin{equation}
{C_{DFT}} = \frac{1}{{\sqrt 3 }}\left( {\begin{array}{*{20}{c}}
1&1&1\\
1&{{e^{\frac{{2\pi i}}{3}}}}&{{e^{\frac{{4\pi i}}{3}}}}\\
1&{{e^{\frac{{4\pi i}}{3}}}}&{{e^{\frac{{2\pi i}}{3}}}}
\end{array}} \right).
\end{equation}
If a DFT coin is used instead of the Grover coin on the Cayley graph of the dihedral group, this keeps  a noticeable peak arises around the starting vertex and its reflection vertex in the picture when the initial coin states are ${\left| 1 \right\rangle }$ and ${\left| 2 \right\rangle }$. Fig.\ref{fig:4} illustrates the different results of probability distribution produced from the same initial states as  Fig.\ref{fig:3}. From Fig.\ref{fig:4}, we see that though the probability distribution of three-state DFT DTQW exists peaks when the initial coin states are ${\left| 1 \right\rangle }$ and ${\left| 2 \right\rangle }$, the values are lower. And that there is a shortfall of central peaks when the initial coin states are ${\left| 0 \right\rangle }$ and $\frac{1}{{\sqrt 3 }}(\left| 0 \right\rangle  + \left| 1 \right\rangle  + \left| 2 \right\rangle )$.

 Localization is a common feature for three-state DTQW, but it does not necessarily occur for all three-state DTQW. We can observe localization around the starting vertex and its reflection vertex in the Fig.\ref{fig:3}, while the Fig.\ref{fig:4}(a) and (d)  do not possess the localization property. Our results tell us which kind of coin and initial coin states we should choose if we want to build a three-state DTQW with or without localization. For the suitable problem, all kinds of coins operator may find its place in a useful quantum walk algorithm with right initial coin state.

\begin{figure}
\centering
\subfigure[]{
\includegraphics[height=3.5cm]{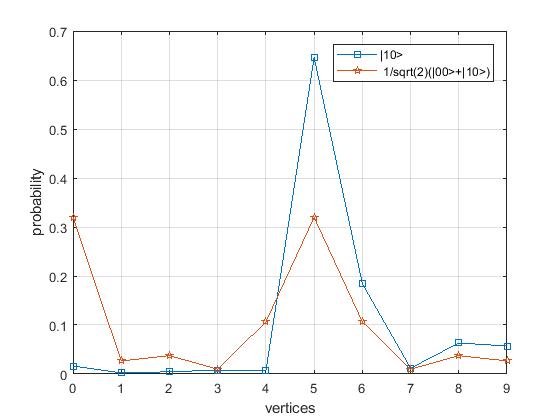}

}
\subfigure[]{
\includegraphics[height=3.5cm]{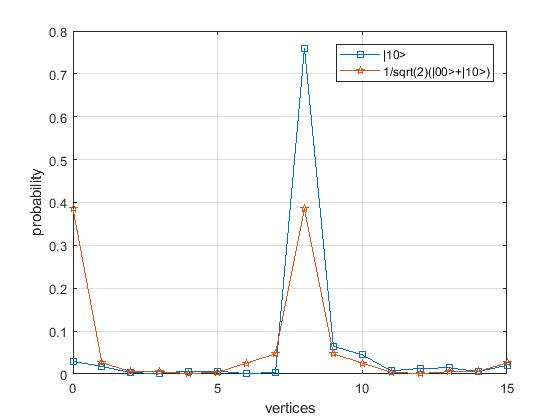}
}
\subfigure[]{
\includegraphics[height=3.5cm]{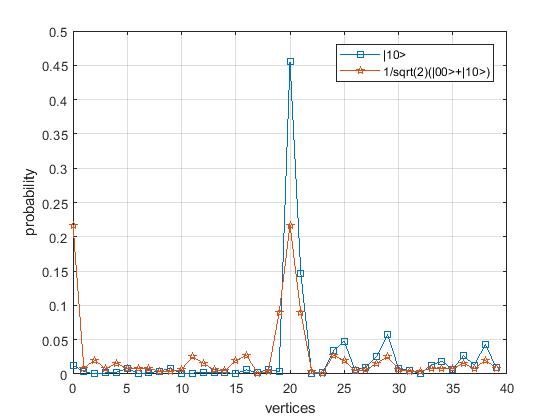}
}
\subfigure[]{
\includegraphics[height=3.5cm]{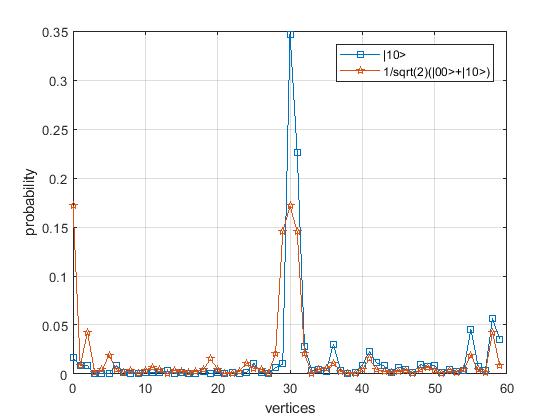}
}
\renewcommand{\figurename}{Fig.}
\caption{Probability distribution of the three-state DTQW with Grover coin on the Cayley graph of the dihedral group $D_N$ with $N = 5(a), 8(b), 20(c) and 30(d)$. The  initial position states are ${\left| 0 0 \right\rangle _V}+{\left| 1 0 \right\rangle _V}$ (pentagram) and ${\left| 10 \right\rangle _V}$ (square), the initial coin state is $\frac{1}{{\sqrt 3 }}(\left| 0 \right\rangle  + \left| 1 \right\rangle  + \left| 2 \right\rangle )$.}
\label{fig:5}
\end{figure}

In order to illustrate the influence of the initial position state on the resulting distribution, two initial states were used to obtain these plots. We choose the initial position states  ${\left| 0 0 \right\rangle _V}+{\left| 1 0 \right\rangle _V}$ and ${\left| 10 \right\rangle _V}$, the initial coin state is  $\frac{1}{{\sqrt 3 }}(\left| 0 \right\rangle  + \left| 1 \right\rangle  + \left| 2 \right\rangle )$. In the case of the  position state is ${\left| 0 0 \right\rangle _V}+{\left| 1 0 \right\rangle _V}$, the resulting probability distribution is symmetric with respect to the starting position. This phenomenon is consistent with  the symmetry of the dihedral group which makes dihedral group widely used in various fields of mathematics, computer science and the natural sciences. In the other case, if the position state is set to ${\left| 10 \right\rangle _V}$, the resulting probability distribution does not have this property, see Fig.\ref{fig:5}. Comparison of the probability distribution for different numbers of vertices is also presented in Fig.\ref{fig:5}. Obviously, the only significant contributions stem from the vertices  is that the probability at the starting vertices is going to be lower as the number of vertices increases.

\subsection{Time-averaged probability}
\label{sec:4.2}
In the previous section, we mainly concentrated on the probability$P({X_t} = X)$ and the long-time limit of the return probability $ \bar P({X_t} = 0)$. And we know that the probability $\mathop {\lim }\limits_{t \to \infty } \bar P({X_t} = 0)$ is only dependent on the initial coin state  $\mathbf{[\alpha, \beta ,\gamma]}^\intercal$ and  the system size $N$. Inui\cite{inui2005one} has shown that the three-state Grover DTQW exhibit localization, but the probability of finding the particle at the original oscillates, because of the existence of degenerate eigenvalues $1$ and $-1$. While the probability of finding the particle can converge in the limit of ${t \to \infty }$, if the degenerate eigenvalue is $1$ only.

\begin{figure}
\centering
\subfigure[]{
\includegraphics[height=4.0cm]{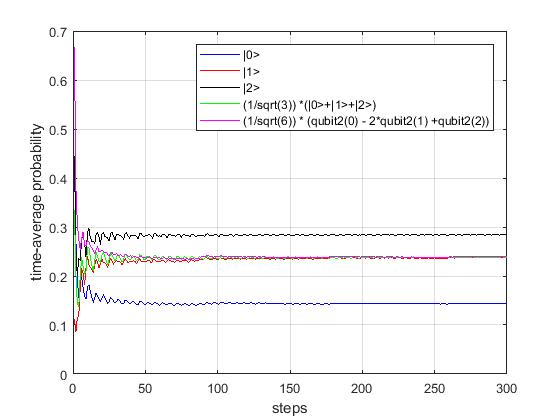}
}
\subfigure[]{
\includegraphics[height=4.0cm]{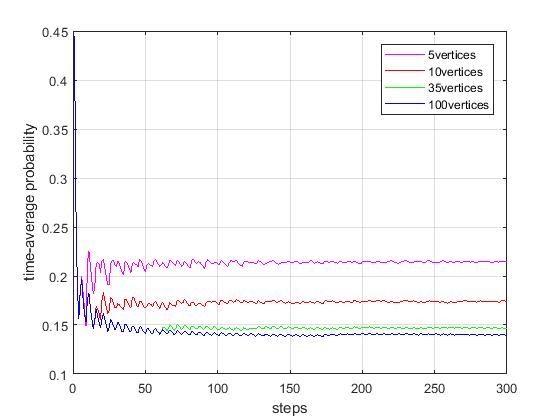}
}
\renewcommand{\figurename}{Fig.}
\caption{(a)Time dependence of the probability of finding a particle at the origin starting from an initial position state ${\left| 10 \right\rangle _V}$ for the three-state Grover DTQW on the Cayley graph of the dihedral group $D_50$ with various initial coin states ${\left| 0 \right\rangle}$ (blue line), ${\left| 1 \right\rangle}$ (red line), ${\left| 2 \right\rangle}$ (black line), $\frac{1}{{\sqrt 3 }}(\left| 0 \right\rangle  + \left| 1 \right\rangle  + \left| 2 \right\rangle )$ (green line), and $\frac{1}{{\sqrt 6 }}(\left| 0 \right\rangle  -2*\left| 1 \right\rangle  + \left| 2 \right\rangle )$ (magenta line).
  (b)Time dependence of the probability of finding a particle at the origin starting from an initial position state ${\left| 10 \right\rangle _V}$ for the three-state Grover DTQW on the Cayley graph of the dihedral group $D_{50}$ with the initial coin state ${\left| 0 \right\rangle}$. The four kinds of initial vertices are 5 vertices(magenta line), 10 vertices(red line), 35 vertices(green line) and 100 vertices(blue line), respectively.}
\label{fig:6}
\end{figure}

We perform numerical simulations after the analytical calculation and the conclusions are summarized in Fig.\ref{fig:6}. The figure manifests the probabilities $\bar P({X_t} = 0)$ oscillate around their corresponding theoretical limiting values $\mathop {\lim }\limits_{t \to \infty } \bar P({X_t} = 0)$. In Fig.\ref{fig:6}(a), the probability decreases quickly near t=0 and fluctuates near a fixed value, when the initial coin state is not really $ \left| 1 \right\rangle$. while the probability increases quickly near t=0, when the initial coin state is $\left| 1 \right\rangle$. Furthermore, we find that probability fluctuates near the same value finally, when the initial coin states contain  $ \left| 1 \right\rangle$. In Fig.\ref{fig:6}(b), it is clearly that, for different system sizes $N$, the probability at the origin oscillates periodically with different patterns. These oscillations clearly exhibit tendencies to converge, which indicate that the walker does have a non-zero probability to be localized. Furthermore, we observe that the larger the $N$ is, the smaller the probabilities is.

\section{Summary}
\label{sec:5}
Quantum walks are one of the elementary techniques of developing quantum algorithms, most of these research findings can be viewed as quantum walks on Abelian groups. However, quantum walks on non-Abelian groups are more suitable for modeling complicated situations than quantum walks on Abelian groups. The dihedral group is one of the simplest non-Abelian groups. Due to its wealth of symmetries, the dihedral group has been studied extensively. They are of particular interest in various fields of mathematics, computer science and the natural sciences. Random walks on groups play an essential role in various fields of natural science, ranging from solid-state physics and polymer chemistry, to mathematics and computer science. Motivated by the immense success of random walk methods in the design of classical algorithms, we pay much attention to DTQW on the Cayley graph of the dihedral group.

Considering the characteristics of the elements in the dihedral group, we propose a model of three-state Grover DTQW on the Caylay graph of the dihedral group and study its basic properties. The walker governed by three-state quantum walk moves to the rotation, the reflection, and stay at the same position. we concentrate on the probability distribution and the long-time limit of the return probability starting from the origin. We calculate the probability of finding the particle at each position after a given number of steps and we provide a formula for the time-averaged limiting probability distribution for the discussed model.

What is more, the analysis of three-state Grover DTQW on  the Caylay graph of the dihedral group clearly demonstrates that  the size of the underlying dihedral group and the coin operator in itself could determine whether localization occur. During our simulation, we find a rare and interesting
result, that time-averaged  probability fluctuates near the same value finally when the initial coin state contains $\left| 1 \right\rangle$. In addition, compared with three-state Grover DTQW with different system sizes $N$, we observed that the smaller the $N$ is, the larger the probabilities  $\mathop {\lim }\limits_{t \to \infty } \bar P({X_t} = 0)$ is, which indicates that a stronger localization around the starting vertex and its reflection vertex can be observed. Localization is a common feature for three-state DTQW, but it does not necessarily occur for all three-state DTQW. For example, the three-state DFT DTQWs do not always possess the localization property. For the right problem, all kinds of coin operator may find its place in a useful quantum walk algorithm with right initial coin state.

Obviously, these conclusions may widen some possible applications of the model. We anticipate that the abundant phenomena of three-state Grover DTQW on the Caylay graph of the dihedral group will be useful in quantum computation and quantum simulation.

\begin{acknowledgements}
This work was supported by the National Natural Science Foundation of China (Grant Nos. 61701229, 61702367,61901218), Natural Science Foundation of Jiangsu Province, China (Grant No. BK20170802, BK20190407), Postdoctoral Science Foundation funded Project of China (Grant Nos. 2018M630557, 2018T110499), Jiangsu Planned Projects for Postdoctoral Research Funds (Grant No. 1701139B), The Open Fund of the State Key Laboratory of Cryptology, China (Grant Nos. MMKFKT201914)
\end{acknowledgements}

\end{document}